\documentclass[cameraready]{Interspeech}

\usepackage{graphicx} 
\usepackage[normalem]{ulem} 

\title{Sub-Model Short-Term Memory Convolutions for Keyword Spotting Systems on Device}

\author[affiliation={1}]{Paweł}{Warlewski}
\author[affiliation={1}]{Artur}{Czeczko}
\author[affiliation={1}]{Artur}{Szumaczuk}
\author[affiliation={2}]{Grzegorz}{Stefański}
\author[affiliation={1}]{Szymon}{Klimaszewski}

\address{
    $^1$ Samsung R\&D Institute Poland\\
    $^2$ Samsung AI Center Warsaw
}

\email{\{p.warlewski2, a.czeczko, a.szumaczuk, g.stefanski, s.klimaszews\}@samsung.com}


\usepackage{comment}

\begin{document}

\maketitle

\begin{abstract}
Keyword Spotting (KWS) is becoming increasingly important as voice-controlled devices grow more widespread. While voice interaction with smartphones and smart TVs is already common, deploying KWS on heavily resource-constrained edge devices such as wearables remains challenging. These systems must meet high accuracy requirements while operating under strict constraints on computational power, memory footprint, and real-time latency. In this work, we present an application of the STMC (Short-Term Memory Convolutions) framework to adapt a modular CNN model for online, LSTM-like inference. Our approach reduces power consumption and redundant computations while maintaining the stability and simplicity of training CNNs. We achieve up to 82\% and 46\% MCPS reduction compared to equivalently frequent standard CNN execution and vanilla STMC, respectively. The best configuration achieves 93.8\% accuracy on the 11-class Google Speech Commands task and 97.1\% on the same task with zero-padded data.
\end{abstract}

\section{Introduction}

Keyword Spotting (KWS) is a well-defined task of detecting a phrase or a set of phrases within the incoming audio stream. The task is becoming increasingly popular thanks to the advent of voice-controlled devices, e.g., TVs, smartphones, etc. The problem becomes progressively more difficult when available resources such as computational power and memory become scarce, as in the case of wearable devices. To tackle the heavily resource-constrained KWS task, the researchers have widely adopted CNNs, which are known for their small memory footprint and proven performance across various tasks \cite{zhang2018helloedgekeywordspotting}. A variety of architectures have been proposed, including ResNet-based models \cite{small-cnn-kws-3}, custom strided CNNs \cite{small-cnn-kws-2}, MatchboxNet \cite{Majumdar_2020}, depthwise separable CNNs \cite{Xu_2020, Bartoli_2025}, or even small MLP networks \cite{small-cnn-kws-1}. Recently, the KWS task has also been tackled using transformers \cite{transformer-kws-1, transformer-kws-2}, but high computational complexity and memory footprints make them inadequate for resource-constrained environments, such as wireless earbuds or smartwatches. Recent TinyML efforts, such as MCUNet \cite{lin2020mcunettinydeeplearning}, demonstrate the feasibility of deep learning on microcontrollers under strict memory constraints.

A practical KWS system must perform inference with high temporal granularity on a continuous audio stream since the target phrases can be very short and may occur at any time in the incoming signal. Ensuring a small memory footprint using a vanilla CNN requires running the model in a sliding-window fashion with a small hop for high granularity. This setting introduces a lot of redundant calculations and may violate the computational power restriction. Recently, the Short-Term Memory Convolutions (STMC) framework \cite{GS-STMC} was proposed to tackle the issue by buffering intermediate results and reusing them when necessary. A slight increase in the memory footprint allowed the reduction of computational overhead and the transformation of the CNN into an LSTM-like model. Thanks to additional memory, the STMC CNNs can process input in chunks much smaller than the receptive field, which reduces the size of the required input buffer to balance the memory footprint. Since fully sequential input processing is not required for KWS, further optimizations and footprint reductions are possible.   

The contributions of this paper are as follows: (1) Propose a small-footprint CNN architecture that achieves good performance on well-established data. (2) Adopt the CNN architecture to the online keyword spotting task via extending and optimizing the STMC scheme to further reduce model footprint while preserving accuracy.
\section{Method}

This section introduces SM-STMC (Sub-Model Short-Term Memory Convolutions), an optimization of STMC that reduces redundant computations in online convolutional networks.

\subsection{State Redundancy in the Classical Approach}

STMC introduces temporal memory buffers for each convolutional and pooling layer. This mechanism caches intermediate activations from previous time steps and reuses them in subsequent computations, enabling online CNN inference with negligible computational overhead. The original motivation for this design was frame-synchronous tasks such as speech enhancement, where producing an output for every incoming frame is both necessary and desirable.

However, when STMC is applied to an architecture containing pooling layers, the buffering strategy leads to additional temporal states whose usefulness depends on the target task. In particular, a max-pooling operation with a stride greater than one increases the number of internal temporal states by a factor equal to the stride. STMC is explicitly designed so that each incoming input frame yields a valid network output, and all states produced by the buffering mechanism therefore correspond to well-defined CNN computations.

For classification-style problems, or other tasks where an output update is not required at every frame, many of these intermediate temporal states become unnecessary. Although they contain valid information and correctly propagate pooled activations over time, they do not provide any additional benefit when predictions are only consumed at a lower temporal resolution. In such settings, maintaining all pooled states introduces redundancy from the perspective of the downstream task.

\begin{figure}[th]
    \centering
    \includegraphics[width=0.5\linewidth]{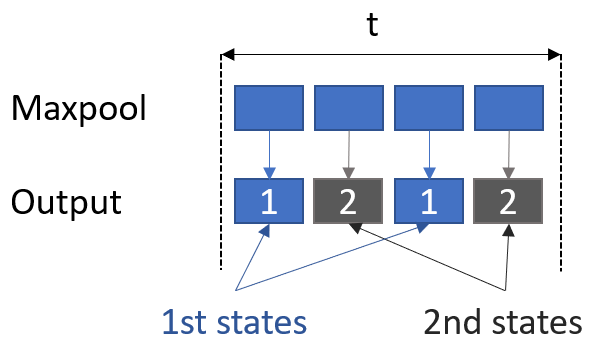}
    \caption{New states after pooling layer with the stride of 2.}
    \label{fig:states_after_pool}
\end{figure}

This behavior is illustrated in Figure~\ref{fig:states_after_pool}, where a pooling layer with a stride of 2 produces two alternating temporal states. Both states are required to preserve the correct temporal evolution of features, but only one may be relevant for producing a classification decision at a given evaluation point.

More generally, for a network with $N$ pooling layers of stride $S$, STMC cycles through $S^N$ internal temporal configurations. All of these states correspond to valid network evaluations and are necessary to support frame-level output generation. Nevertheless, in tasks where outputs are only required sporadically (e.g., sequence-level classification), many of these states do not contribute any additional useful information to the final decision and can be viewed as redundant from a task-driven perspective.

\subsection{Sub-model Decomposition and State Scheduling}

SM-STMC reduces state redundancy by decomposing the convolutional backbone into a sequence of sub-models, each ending with either a pooling layer or an original output layer. Each sub-model corresponds to an adequate convolutional block and is executed independently.

These sub-models do not share layers or weights. Instead, they are connected through memory buffers, as in the original formulation. This decomposition preserves the original data flow and, if the sub-models are executed consecutively, the original network structure.

Given the architecture of the model, we can schedule the order in which the sub-models are executed. Let $N$ denote the number of pooling layers plus the output layer. At time step $t$, SM-STMC selects the deepest sub-model that should be executed at the current iteration. Let $n \in \{0, 1, \dots, N-1\}$. The index of the final sub-model executed at iteration $t$ is given by:

\begin{equation}
    n^*=\max \Bigl\{ n \mid t \bmod 2^n = 2^n -1 \Bigl\}
    \label{equation:scheduled_models}
\end{equation}

This approach is conceptually related to early-exit methods (e.g., BranchyNet \cite{teerapittayanon2017branchynetfastinferenceearly}), in which the execution path is dictated by the current internal state.

\begin{figure}[th]
    \centering
    \includegraphics[width=1.0\linewidth]{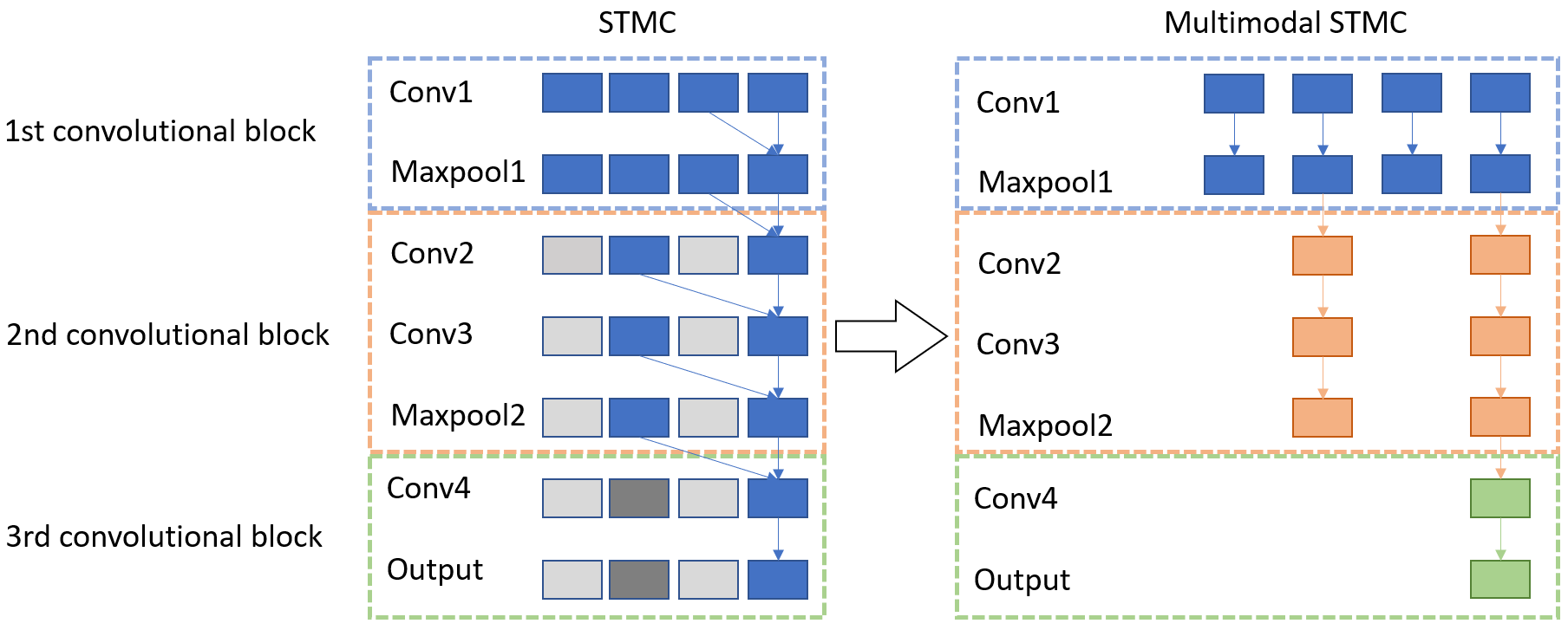}
    \caption{Difference between STMC (left) and SM-STMC (right).}
    \label{fig:stmc_v_mmstmc}
\end{figure}

The right side of Figure~\ref{fig:stmc_v_mmstmc} illustrates this process for a network with three convolutional blocks.

In standard STMC (left side of Figure~\ref{fig:stmc_v_mmstmc}), all blocks are active at every time step. In SM-STMC (right side), the execution schedule becomes: (1) block 1, (2) block 1 $\rightarrow$ block 2, (3) block 1, (4) block 1 $\rightarrow$ block 2 $\rightarrow$ block 3.

\subsection{Memory Management and Computational Complexity}

SM-STMC uses the same memory management principles as traditional STMC, but excludes memory associated with redundant states. Execution of the sub-model alters only its corresponding memory buffers. With this scheme, memory is allocated and accessed as if only a single temporal state were maintained.

This approach utilizes the whole network only once every $2^N$ steps. Intermediate steps execute fractions of the total computation, substantially reducing the workload. As a result, the average number of executed layers per time step is reduced, leading to a proportional reduction in operations. This reduction scales with network depth, while the latency and alignment remain unchanged.

\subsection{Training and Deployment}

STMC and SM-STMC do not require re-training. The convolutional backbone is trained independently using standard methods. During evaluation, STMC is applied as a deterministic temporal extension of the trained network. This approach does not incorporate any additional weights or parameters. 

SM-STMC is implemented using multiple independent sub-models rather than dynamic conditional execution within a single model. This design avoids runtime control flow, which is unsupported by common deployment frameworks such as TensorFlow Lite. As a result, the method ensures compatibility with standard conversion pipelines and enables efficient deployment, provided that memory management and scheduling are implemented on the target device.

\section{Experiments}

This section evaluates the proposed SM-STMC method in terms of recognition performance and computational efficiency. We compare four approaches: (1) a standard VGG-based (Visual Geometry Group) \cite{simonyan2015deepconvolutionalnetworkslargescale} acoustic embedding with an MLP classifier, (2) the same VGG model evaluated eight times per second using a sliding-window scheme, (3) standard STMC with the same MLP classifier, and (4) the proposed SM-STMC with the same classifier.

In all cases, except the first, we evaluated the classifier eight times per second. With the STMC-based approach, we can invoke the classifier after every time bin, but this is not required. Instead, we did it every eight frames, resulting in 7.8125 inferences per second.

All evaluations were conducted on two different models with different numbers of states. The results were also compared with a conventional LSTM network.

\subsection{Dataset and Experimental Setup}

Training and evaluation were conducted on the Google Speech Commands (GSC) \cite{GSC} dataset, with 10 target classes (``Yes'', ``No'', ``Up'', ``Down'', ``Left'', ``Right'', ``On'', ``Off', ``Stop'', and ``Go''), the rest of the classes served as negative class for the contrast. Two evaluation sets were used: (1) the standard GSC test dataset, where each sample was 1 second long, and (2) a modified dataset where 0.5 seconds of silence was added to both the beginning and the end of each sample, resulting in 2 seconds of audio. The purpose of the second dataset is to showcase the limitations of the offline approach. In such cases, the target word may be only partially contained within the input window, resulting in incorrect negative prediction.

The first VGG backbone model has a receptive field of 41 temporal frames, and the second has a receptive field of 38. Input audio is preprocessed into a Mel spectrogram using a window size of 1024 samples and a hop size of 256 samples, resulting in 62.5 temporal bins per second. Frames corresponding to one second of audio are then fed as an input to the model. In two seconds, this would result in two full input buffers.

The VGG backbone was trained offline using a standard supervised training scheme. STMC and SM-STMC are inference-time techniques that require no additional retraining, so the original set of weights was applied. All approaches use the same MLP classifier as well.

To demonstrate the effectiveness of this method in an online setting, we evaluate the STMC classifiers eight times per second. To match the setup for the vanilla VGG model, we evaluated it using a sliding-window scheme at the same temporal rate.

Since STMC and SM-STMC achieve identical recognition performance, we report results for only one of them.

The on-device experiments were run on the ARM Cortex-M55 CPU at 196 MHz. Its crucial feature is M-Profile Vector Extension (MVE), also called ARM Helium. MVE uses 128-bit registers, allowing up to 16 parallel operations in int8 data, the type our models are quantized to using standard integer-only inference techniques \cite{jacob2017quantizationtrainingneuralnetworks}. To further reduce memory consumption, TensorFlow Lite Micro \cite{david2021tensorflowlitemicroembedded} has been chosen as the deployment framework, optimizing MCU inferences using libraries such as CMSIS-NN \cite{lai2018cmsisnnefficientneuralnetwork}.

\subsection{Parameters}

Table~\ref{tab:parameters} summarizes the number of parameters for each model configuration. Embedding consists of multiple convolutional blocks, and the parameters for each individual block are reported. STMC-based methods use additional memory buffers with a size depending on the kernel, filters, and mels. In the standard STMC, these values are also multiplied by the number of states. The size of these temporal memory buffers is reported in Table~\ref{tab:memory}. STMC\textsuperscript{1} has a total of 4 states, while STMC\textsuperscript{2} has a total of 8 states. Superscripts 1 and 2 denote the first and second backbone configurations, respectively.

\begin{table}[th]
    \centering
    \caption{Parameters of the backbone's embedding and classifier.}
    \setlength{\tabcolsep}{5pt}
    \begin{tabular}{lccc}
    \hline
        Model & Emb & Cls & Total \\
        \hline
        SM-STMC\textsuperscript{1} & 513 + 1776 + 6496 & 18443 & 27228 \\
        SM-STMC\textsuperscript{2} & 105 + 816 + 1472 + 12064 & 18443 & 32900 \\
        LSTM & 20992 & 6155 & 27147 \\
    \hline
    \end{tabular}
    \label{tab:parameters}
\end{table}

\begin{table}[th]
    \centering
    \caption{Buffer size (int8).}
    \begin{tabular}{lc}
    \hline
        Model & Total buffer size \\
        \hline
        \hline
        STMC\textsuperscript{1} & 17184 \\
        SM-STMC\textsuperscript{1} & \textbf{9280} \\
        \hline
        STMC\textsuperscript{2} & 21632 \\
        SM-STMC\textsuperscript{2} & \textbf{7520} \\
    \hline
    \end{tabular}
    \label{tab:memory}
\end{table}

\subsection{Evaluation Metrics}

We report weighted average precision, recall, and F1-score. In the single-label multi-class setting considered here, the weighted average recall is equivalent to classification accuracy. Due to space constraints, only average metrics are reported. In all of the streaming models, we classify embeddings every 8th frame, which corresponds to the minimum sequence of frames for SM-STMC\textsuperscript{2}. For consistency, the classifiers of the remaining models are evaluated at the same temporal intervals, resulting in 7.8125 evaluations per second.

\begin{table}[th]
    \centering
    \caption{Weighted average performance comparison on the standard 1~s GSC dataset.}
    \begin{tabular}{lcccc}
    \hline
        Model & Precision & Recall & F1-score \\
        \hline
        \hline
        VGG\textsuperscript{1} ($1\times1\text{ s}$)   & 0.9362 & 0.9278 & 0.9292 \\
        SM-STMC\textsuperscript{1}             & \textbf{0.9494} & \textbf{0.9382} & \textbf{0.9403} \\
        \hline
        VGG\textsuperscript{2} ($1\times1\text{ s}$)   & 0.9159 & 0.9035 & 0.9055 \\
        SM-STMC\textsuperscript{2}             & \textbf{0.9322} & \textbf{0.9162} & \textbf{0.9188} \\
        \hline
        LSTM                                   & 0.9569 & 0.9511 & 0.9521 \\
    \hline
    \end{tabular}
    \label{tab:results_1s}
\end{table}

\begin{table}[th]
    \centering
    \caption{Weighted average performance comparison on the extended 2~s GSC dataset with silence added.}
    \begin{tabular}{lcccc}
    \hline
        Model & Precision & Recall & F1-score \\
        \hline
        \hline
        VGG\textsuperscript{1} ($1\times1\text{ s}$)           & 0.8353 & 0.4981 & 0.5798 \\
        VGG\textsuperscript{1} ($8\times1\text{ s}$)           & 0.9733 & 0.9708 & 0.9712 \\
        SM-STMC\textsuperscript{1}                     & \textbf{0.9738}  & \textbf{0.9710} & \textbf{0.9715} \\
        \hline
        VGG\textsuperscript{2} ($1\times1\text{ s}$)           & 0.7790 & 0.6182 & 0.6386 \\
        VGG\textsuperscript{2} ($8\times1\text{ s}$)           & 0.9586 & 0.9524 & 0.9532 \\
        SM-STMC\textsuperscript{2}                     & \textbf{0.9611}  & \textbf{0.9552} & \textbf{0.9561} \\
        \hline
        LSTM                                           & 0.9396 & 0.9196 & 0.9235 \\
    \hline
    \end{tabular}
    \label{tab:results_2s}
\end{table}

The VGG models achieve accuracy comparable to the STMC-based approaches. However, to obtain identical performance, they would need to be evaluated at every time bin.

\subsection{Results on Standard GSC}

Results on the standard GSC dataset are presented in Table~\ref{tab:results_1s}. The standard VGG model achieves overall lower performance than STMC, with a drop of 1.04\%\textsuperscript{1} in accuracy. This result is expected, since the classifier is invoked more frequently in the online approach. 

\subsection{Results on Extended Inputs}

Table~\ref{tab:results_2s} reports results on the 2-second dataset with silence padding. This evaluation highlights the limitations of offline inference. The standard VGG model failed to correctly classify more than half of the samples. This occurs when a keyword spans the boundary between two input audio windows (e.g. ``do-'' in the first window and ``-wn'' in the second), leading to misclassification. 

\begin{figure}[th]
    \centering
    \includegraphics[width=1.0\linewidth]{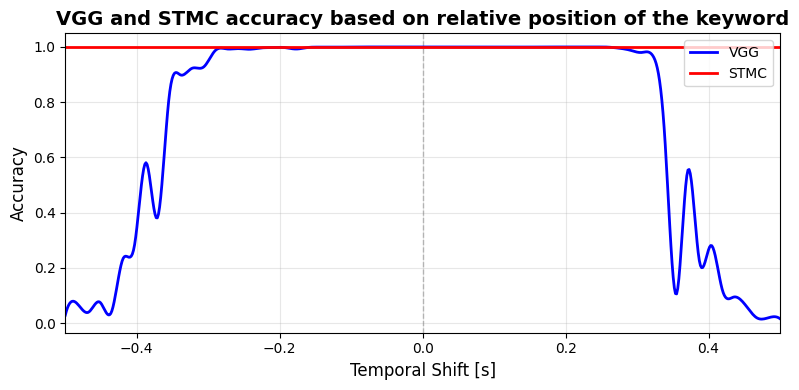}
    \caption{Influence of temporal shift on accuracy. VGG and STMC accuracy on a 2-second zero-padded ``stop'' sample.}
    \label{fig:temporal_shift}
\end{figure}

Figure~\ref{fig:temporal_shift} illustrates this boundary effect using a zero-padded example evaluated under different temporal shifts. When the keyword is centered within the input frame (temporal shift of 0.0~s), both architectures produce correct predictions. However, when the keyword lies across two frames (shift -0.5~s or 0.5~s), the VGG model's accuracy drops substantially, whereas the STMC-based model remains stable.

The VGG model (evaluated eight times per second) and the SM-STMC-based models achieve significantly better results, with an accuracy improvement of 3.28\%\textsuperscript{1} compared to the 1-second setting. Improvement comes from the fact that the keyword is captured across multiple overlapping windows, increasing the likelihood that at least one window contains a more favorable representation for classification.

\subsection{Computational Cost and Latency}

Computational complexity is measured in \textit{million cycles per second} (MCPS) and is reported separately for embedding computation, classifier evaluation, memory management, and total cost, reported in Table~\ref{tab:mcps}. For SM-STMC, the embedding stage consists of multiple independent sub-models corresponding to different convolutional blocks. 

In the case of VGG, evaluating the model eight times per second consumes proportionally more MCPS. Despite the fact that this approach achieved approximately the same results as STMC on 2 seconds of audio, it required almost four times more computational cycles for the first backbone model and three times more for the second.

Compared to the standard VGG baseline, STMC increases computational cost by a factor of 2.27\textsuperscript{1} for the first model, while providing fully online inference. SM-STMC reduces embedding computation and memory management overhead by removing redundant states, decreasing this factor to only 1.53\textsuperscript{1}. For the second model, this factor is respectively 2.78\textsuperscript{2} and 1.47\textsuperscript{2}.

\begin{table}[th]
    \centering
    \caption{Computational cost comparison of embedding, classifier, and memory management in MCPS.}
    \begin{tabular}{lcccc}
    \hline
        Model & Emb & Cls & Memory & Total \\
        \hline
        \hline
        VGG\textsuperscript{1} ($1\times1\text{ s}$)   & 7.40 & 0.02 & -- & 7.42 \\
        VGG\textsuperscript{1} ($8\times1\text{ s}$)   & 59.20 & 0.16 & -- & 59.36 \\
        STMC\textsuperscript{1}                & 16.12 & 0.16 & 0.56 & 16.84 \\
        SM-STMC\textsuperscript{1}             & 10.82 & 0.16 & 0.39 & 11.37 \\
        \hline
        VGG\textsuperscript{2} ($1\times1\text{ s}$)   & 3.66 & 0.02 & -- & 3.68 \\
        VGG\textsuperscript{2} ($8\times1\text{ s}$)   & 29.28 & 0.16 & -- & 29.44 \\
        STMC\textsuperscript{2}                & 9.42 & 0.16 & 0.44 & 10.02 \\
        SM-STMC\textsuperscript{2}             & 5.12 & 0.16 & 0.16 & 5.44 \\
        \hline
        LSTM                                   & 10.29 & 0.05 & 0.01 & 10.35 \\
    \hline
    \end{tabular}
    \label{tab:mcps}
\end{table}

Total detection latency consists of computational delay and the temporal alignment of the keyword within the input buffer. In a streaming context, predictions are generated every 16~ms. To account for the stochastic nature of keyword onset, we report the average latency as the midpoint between the best-case (keyword ends exactly at a buffer boundary) and worst-case (keyword ends just after a boundary) scenarios, plus the specific computational overhead of the model. Latency is measured according to each model's native classification rate, e.g., LSTM every frame, while SM-STMC\textsuperscript{1} every 4th frame.

For the baseline VGG models, classification occurs once per second, leading to a high average latency of approximately 550--600~ms. In contrast, STMC and LSTM models provide frame-synchronous predictions (16~ms granularity), resulting in early detection with "negative" reported latency relative to the keyword offset. This is a consequence of the models capturing the discriminative features of the keyword before the entire keyword has been observed.

\begin{table}[th]
    \centering
    \caption{Computational latency and average latency of the models measured from the end of the uttered keyword to its first corresponding prediction.}
    \begin{tabular}{lcc}
    \hline
        Model & Comp. Latency [ms] & Avg Latency [ms] \\
        \hline
        \hline
        VGG\textsuperscript{1}                 & 101 & 601 $\pm$ 500 \\
        STMC\textsuperscript{1}                & 4 & -156 $\pm$ 8 \\
        SM-STMC\textsuperscript{1}             & 10 & -146 $\pm$ 32 \\
        \hline
        VGG\textsuperscript{2}                 & 50 & 550 $\pm$ 500 \\
        STMC\textsuperscript{2}                & 2 & -158 $\pm$ 8 \\
        SM-STMC\textsuperscript{2}             & 9 & -116 $\pm$ 64 \\
        \hline
        LSTM                                   & 5 & -167 $\pm$ 8 \\
    \hline
    \end{tabular}
    \label{tab:latency}
\end{table}

The SM-STMC approach introduces a controlled trade-off: by classifying every 4th frame (SM-STMC\textsuperscript{1}) or every 8th frame (SM-STMC\textsuperscript{2}), the average latency increases compared to vanilla STMC able to classify data every frame, but remains within the margin of classification time. Furthermore, when comparing the time required to process an equivalent duration of audio (computationally-wise), SM-STMC\textsuperscript{2} is 1.78 and 4.44 times more efficient than STMC and LSTM, respectively.

\section{Conclusion}

In this work, we proposed SM-STMC, a sub-model decomposition and scheduling strategy that reduces redundant temporal states in online convolutional networks. The method extends the original STMC framework by exploiting the task-level requirement of sparse output generation, thereby eliminating unnecessary internal states by using model decomposition and executing only the required network depth with static scheduling.

Experimental results on the Google Speech Commands dataset demonstrate that SM-STMC preserves the high recognition performance of CNNs while achieving substantial efficiency gains. Specifically, we observed an MCPS reduction of up to 82\% compared to the standard sliding-window baseline and a memory footprint reduction of 65\% compared to the original STMC approach. While the reduced classification frequency introduces a marginal increase in latency, the resulting values remain well within the requirements for real-time human-computer interaction.

Importantly, SM-STMC requires no retraining and introduces no additional parameters, making it fully compatible with existing pretrained models. The use of static scheduling and independent sub-models ensures that the system can be deployed directly via standard frameworks like TensorFlow Lite on resource-constrained edge devices, such as wearables and hearables. 

\section{Generative AI Use Disclosure}
LLM (Large Language Model) has been used to polish and to find grammar errors in this manuscript. However, it was not used to directly write it or to generate code used in this work.

\bibliographystyle{IEEEtran}
\bibliography{bibliography}

\end{document}